\documentclass[a4paper]{llncs}

\usepackage{alltt}
\usepackage{amssymb}
\usepackage{caption}
\usepackage{paralist}
\usepackage{nicefrac}
\usepackage{booktabs}
\usepackage{xspace}
\usepackage{amsmath}
\usepackage{marvosym}
\usepackage{wasysym}
\usepackage{verbatim}
\usepackage{float}
\usepackage{hyperref}
\usepackage{multirow}
\usepackage{cite}
\usepackage[capitalize]{cleveref}
\usepackage[per-mode=symbol,detect-all]{siunitx}
\usepackage{graphics}
\usepackage{amssymb}
\usepackage{enumitem}
\usepackage[table]{xcolor}
\usepackage{booktabs}
\usepackage{colortbl}
\usepackage{ifthen}
\usepackage{mathdots}
\usepackage{arydshln}
\usepackage[outline]{contour}
\usepackage{graphicx}
\usepackage{wrapfig}
\usepackage{subfig}

\definecolor{navy}{RGB}{0,0,128}

\newcommand{\R}{\mathbb{R}}

\newtheorem{newDefinition}{\textbf{Definition}}

\usepackage{tikz,ifthen,pgfplots}
\usetikzlibrary{arrows,trees,backgrounds,automata,shapes,decorations,plotmarks,fit,calc,positioning,shadows,chains}
\usetikzlibrary{quotes,arrows.meta}
\tikzstyle{every pin edge}=[<-,shorten <=1pt]
\tikzstyle{neuron}=[circle,fill=black!25,minimum size=17pt,inner sep=0pt]
\tikzstyle{input neuron}=[neuron, fill=green!50]
\tikzstyle{output neuron}=[neuron, fill=red!50]
\tikzstyle{hidden neuron}=[neuron, fill=blue!50]
\tikzstyle{small neuron}        =[hidden neuron, draw, minimum size=15pt]
\tikzstyle{small input neuron}  =[input neuron , draw, minimum size=15pt]
\tikzstyle{small output neuron} =[output neuron, draw, minimum size=15pt]
\tikzstyle{annot} = [text width=4em, text centered]
\tikzstyle{nnedge} = [-{stealth},shorten >=0.1cm, shorten <=0.05cm,line width=0.8pt,black]
\tikzstyle{edge} = [->,line width = 0.3pt, shorten >=0.2cm]
\tikzstyle{edgeWide} = [->,line width = 2pt, , shorten >=0.2cm]
\usetikzlibrary{calc}

\tikzset{every picture/.style={line width=0.75pt}} 
\tikzstyle{BadSquare}=[rectangle,fill=red!30!white,minimum size=25pt,inner 
sep=0pt]
\tikzstyle{InitSquare}=[rectangle,fill=green!30!white,minimum size=25pt,inner 
sep=0pt]

\newcommand{\mysubsection}[1]{\medskip\noindent\textbf{#1}}

\newcommand{\relu}{\text{ReLU}\xspace}

\newcommand{\sat}{\texttt{SAT}\xspace}
\newcommand{\unsat}{\texttt{UNSAT}\xspace}

\newcommand{\marabou}{\textit{Marabou}\xspace}

\usepackage{bussproofs}
\usepackage{gensymb}

\newif\ifcomments
\commentstrue

\newif\ifoutline
\outlinefalse

\newif\iflong
\longtrue

\renewcommand{\paragraph}[1]{\vspace{1mm}\noindent{\bf #1}\ }

\begin{document}
	
	\title{veriFIRE: Verifying an Industrial, Learning-Based Wildfire
          Detection System}
	
	\author{
		Guy Amir\inst{1} \and
		Ziv Freund\inst{2} \and
		Guy Katz\inst{1} \and
		Elad Mandelbaum\inst{2} \and
		Idan Refaeli\inst{1}
	\thanks{All authors contributed equally.}
	}
	\institute{
          The Hebrew University of Jerusalem, Jerusalem, Israel\\
           \email{ \{guyam, guykatz, idan0610\}@cs.huji.ac.il}
          \\
          \and
        Elbit Systems --- EW $\&$ SIGINT --- Elisra  Ltd., Holon, Israel\\
       \email{ \{ziv.freund, elad.mandelbaum\}@elbitsystems.com}
           \\
	}
	
	\maketitle
	
	\begin{abstract} 
          In this short paper, we present our ongoing work on the
          veriFIRE project --- a collaboration between industry and
          academia, aimed at using verification for increasing the
          reliability of a real-world, safety-critical system. The
          system we target is an airborne platform for wildfire
          detection, which incorporates two deep neural networks. We
          describe the system and its properties of interest, and
          discuss our attempts to verify the system's consistency,
          i.e., its ability to continue and correctly classify a given
          input, even if the wildfire it describes increases in
          intensity.
            We regard this work as a step towards
          the incorporation of academic-oriented verification tools
          into real-world systems of interest.
		
	\end{abstract}

\section{Introduction}
\label{sec:Introduction}

In recent years, \emph{deep neural networks} (DNNs)~\cite{GoBeCo16}
have achieved unprecedented results in a variety of fields, such as
image recognition~\cite{SiZi14}, speech analysis~\cite{NaShAtAzSh19},
and many others~\cite{SiHuMaGuSiVaScAnPaLaDi16,
  BoDeDwFiFlGoJaMoMuZhZhZhZi16, MnKaSiGrAnWiRi13, LeMoSuSa20,
  JuEvPrGrFiRoTuBaZiPo21}. This success has led to the integration
of DNNs in various safety-critical systems~\cite{DoWaAb21}.

A particular safety-critical application of DNNs is within
\textit{wildfire detection} systems~\cite{ZhXuXuGu16, LiZh20,
  LeKiLeLeCh17, ShGrGoFi17}, whose goal is to detect and alert first
responders to situations that could later become life threatening.
One such airborne system, which is currently being considered by Elbit
Systems for use on aerial vehicles, is based on Infra-Red (IR) sensors that
feed their inputs, usually a series of image frames, to multiple
neural networks --- which then determine whether the images contain a
wildfire. Naturally, it is
possible that (a) the system will mistakenly issue an alert when a
wildfire does not exist, or, worse, that (b) the system will fail to
issue an alert when the images do indicate the existence of a
wildfire. The second kind of failure is clearly very dangerous, and
could potentially jeopardize human lives. Consequently, potential users of
the system require it to be extremely reliable.

Although DNN-based systems are highly successful, prior research has
shown that even complex and highly-accurate DNNs are prone to
errors. For example, small input perturbations, due to
either random noise or adversarial attacks, are known to cause modern
DNNs to fail miserably~\cite{MoSeFaPa17, KuGoBe18, GoShSz14}. Such
issues raise serious concerns regarding the trustworthiness of a
DNN-based wildfire detection system, and could delay or prevent its
deployment.

In order to address such issues and facilitate the certification of
DNNs, the formal methods community has recently suggested various
tools and approaches for \emph{formally verifying} the correctness of
DNNs~\cite{KaBaDiJuKo17, GeMiDrTsChVe18, WaPeWhYaJa18, LyKoKoWoLiDa20,
  HuKwWaWu17, ZhShGuGuLeNa20, KoLoJaBl20, BaShShMeSa19, LoMa17,
  KaBaDiJuKo21, StWuZeJuKaBaKo21, IvCaWeAlPaLe20, DuJhSaTi18,
  ZeWuBaKa22, JiTiZhWeZh22}, based on reachability analysis and
abstract interpretation~\cite{ GeMiDrTsChVe18, TjXiTe17, LoMa17},
SMT-solving~\cite{Eh17, KaBaDiJuKo17, KaHuIbHuLaLiShThWuZeDiKoBa19,
  HuKwWaWu17, KuKaGoJuBaKo18, GoKaPaBa18, AmWuBaKa21}, and other
methods. Given a DNN and a specification, these techniques allow us to
formally prove that the DNN satisfies the specification for \emph{any}
possible input of interest (see Appendix~\ref{sec:appendix:Background}
for additional details). However, despite the rapid improvement in DNN
verification technology, there remains a gap between the capabilities
of verification tools developed by academia, and the actual needs of
industrial teams.  First, academic tools often face scalability
issues, and may be unsuitable for verifying industrial-sized DNNs with
millions of neurons. Second, academic-oriented verification tools may
not support the various DNN specifications used in
industry. Consequently, practitioners often resort to using various
forms of testing, and not verification, when attempting to certify
real-world DNNs.

	
In this paper, we describe our ongoing work on the \emph{veriFIRE}
project --- a collaboration between Elbit Systems and the Hebrew University,
aimed at formally verifying the correctness of the aforementioned
wildfire detection system. As part of this project, our goals are to
(1) produce formal specifications for this system, which could then be
formulated into DNN verification tools; and (2) enhance and extend
existing verification technology, so that it can be successfully
applied to this system.


\section{The veriFIRE Project}
\label{sec:veriFIRE}

\begin{figure}[htp]
  \begin{center}
   \includegraphics[width=0.8\linewidth]{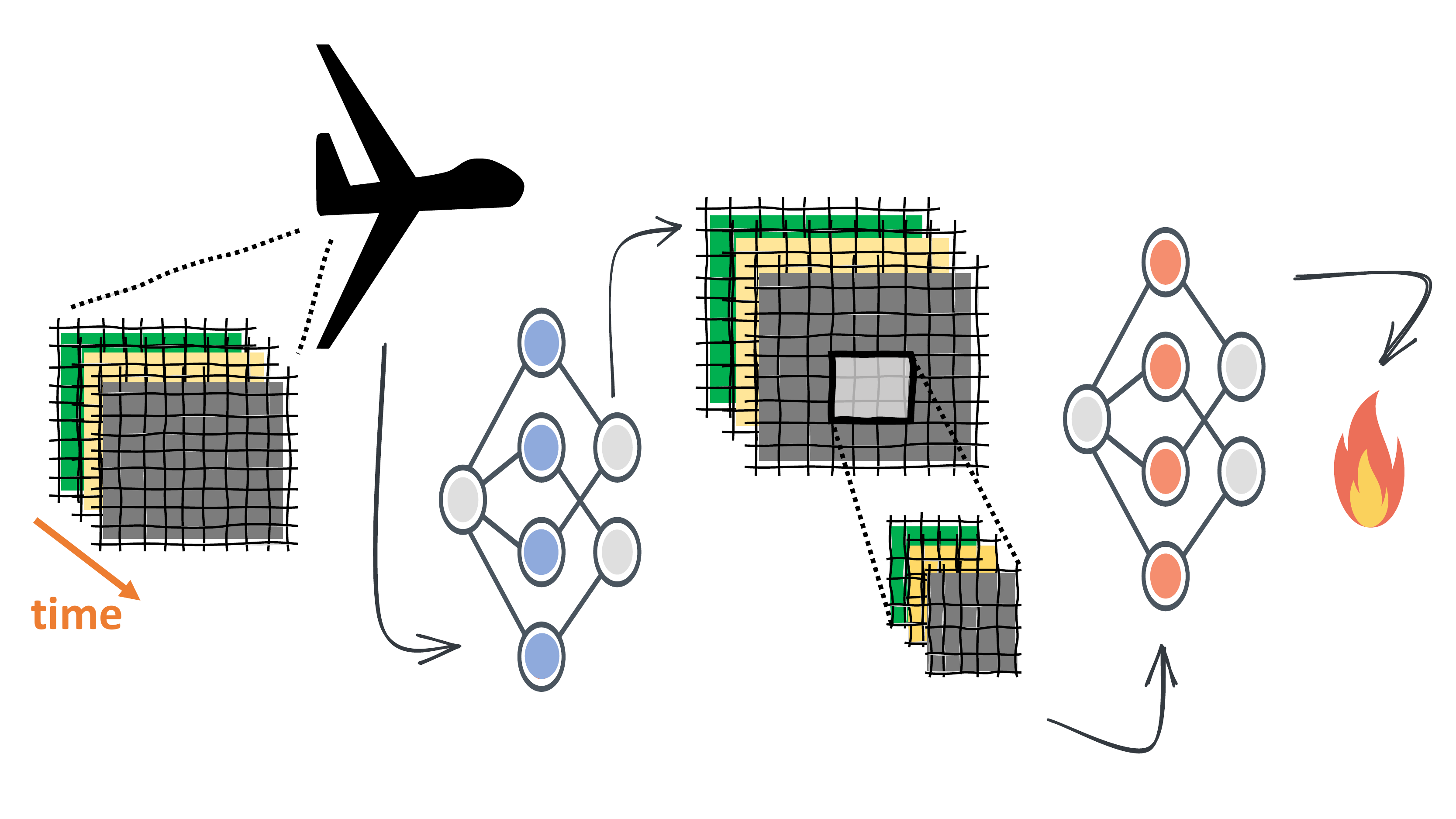}
  \end{center}
  \caption{A scheme of the airborne wildfire detection system. At
    first, an airborne platform takes multiple IR images, and uses the
    first DNN to detect candidate areas, in which a wildfire is
    suspected. Next, these candidates are passed to a second DNN, which
    determines whether a wildfire has truly occurred, or not. }
	\label{fig:classificationScheme}
\end{figure}

\mysubsection{The Platform.}  The veriFIRE project is a recent
and ongoing collaboration between Elbit Systems and the Hebrew
University. It involves an airborne wildfire detection system,
designed to be mounted on aerial vehicles (AVs) --- from small drones,
to large manned or unmanned aircraft --- being
manufactured by Elbit Systems (see Fig.~\ref{fig:classificationScheme}). The
airborne system consists of the following components:
\begin{inparaenum}[(i)]
\item a set of infra-red (IR) sensors, located at different spots on
  the AV, and pointing at different angles.  These sensors produce temporal 
  image streams of the \emph{background} surrounding the AV;
\item a first, convolutional DNN, which receives the image streams
  generated by the IR sensors, and produces candidate detections, based on
  temporal changes as detected when compared to previous images of the background. Each
  candidate detection is a stream of slices (through time) taken from the background image streams,
  around the suspicious areas; and
\item a second convolutional DNN, which receives a candidate detection, produced by
  the first DNN, and determines whether it is a wildfire (at its early
  stages), or a false detection of the first DNN.
\end{inparaenum}
The goal of the veriFIRE project is to ensure the overall reliability
of the system, by verifying the correctness of its DNN components.

Training the wildfire detection platform is performed using a
proprietary simulator that automatically generates synthetic images,
by adding simulated wildfire images to recorded background images. Given two datasets,
one containing only normalized wildfire signals ($\mathcal{S}$) with
no background, and another for background images ($\mathcal{B}$) which
do not contain any wildfires, the simulator creates a new dataset of
synthetic images, each one generated by combining a wildfire image
with a background image, in a process referred to as
\emph{planting}. More formally, for any $x_s \in \mathcal{S}$,
$x_b \in \mathcal{B}$, the simulator uses a planting function $p$ to
produce a realistic image $I = p(\epsilon \cdot x_s, x_b)$, which
contains the wildfire with intensity $\epsilon$. At its early stages, a wildfire is a sub-pixel in the sensor's field of view, and thus the planting function can be treated as a linear combination of the wildfire image and the background image. We note that this
methodology is common practice, and is acceptable to Elbit Systems' clients.

Although the dataset is large enough to produce sufficiently many test
samples, statistical testing alone is inadequate for guaranteeing the
platform's reliability. Specifically, clients may wish to guarantee
that some performance features are not random --- for example, it is
required that if a small wildfire is detected by the platform in a
given scenario, a stronger wildfire will definitely be detected as
well. Thus, we began by focusing on formally verifying the correctness
of the second DNN used, which we term $N$.  This network can be
regarded as a mapping $N : \mathbb{R}^{n\times k} \rightarrow \R$,
where $n$ is the number of pixels in each image, and $k$ is the number
of time-steps observed. When presented with a stream of input images
$x \in \mathbb{R}^{n\times k}$, $N$ computes a score, $N(x)$; and if
this score exceeds a threshold $\delta$, then $N$ classifies $x$ as an image containing a wildfire. The value of $\delta$ is determined according to the clients' needs, as a balancing point between the empirical false-alarm rate and its tradeoff with the empirical positive-detection rate, after a short evaluation period.
The network $N$ is comprised of three convolution layers~\cite{SiZi14,
  KrSuHi12}, each one followed by a max-pooling layer and two
fully-connected layers. In the last layer, the network has a single output node with a
sigmoid activation, which serves as the output of the entire DNN.


  



\mysubsection{Consistency.} One main challenge in the veriFIRE project
is to produce formal specifications for $N$. Ideally, we would like to
prove that $N$ correctly identifies any possible wildfire within any
possible image, but this is difficult to formulate rigorously.
Current state-of-the-art verification tools focus primarily on
verifying local adversarial robustness~\cite{TjXiTe17, GoKaPaBa18,
  WeZhChSoHsBoDhDa18, GeMiDrTsChVe18, LyKoKoWoLiDa20, LeKa21,
  OsBaKa22, CaKoDaKoKaAmRe22}, i.e., on proving that a DNN continues
to correctly classify an input in the presence of slight
perturbations; but we have observed that this kind of property is of
limited interest to potential clients of the system.  Thus, a new kind
of specification is required for this process. With that in mind, we
introduce the definition for \emph{local consistency}:


\indent
\begin{newDefinition}[\textbf{Local Consistency}]
  Given a deep neural network
  $N : \mathbb{R}^{n\times k} \rightarrow \mathbb{R}$, a wildfire signal image stream $x_s \in \mathcal{S} $, and an input background image stream
  $x_b \in \mathcal{B} $, we
  say that $N$ is $(x_s,x_b)$-locally-consistent if for every
  $\epsilon_1 \geq \epsilon_2$, it holds that $ N(p(\epsilon_1 \cdot x_s,x_b)) \geq N(p(\epsilon_2 \cdot x_s,x_b)) $,
  where $p : \mathbb{R}^{n\times k} \times \mathbb{R}^{n\times k} \rightarrow \mathbb{R}^{n\times k}$ is a planting function, such that $p(s, b)$ plants the signal $s$ into the background $b$.
\end{newDefinition}

Intuitively, local consistency in this context means that if the
original image $x$ was determined to contain a wildfire (i.e., $N(x)$
exceeded the threshold $\delta$), then any image stream with a
\emph{stronger signal}, e.g., a larger wildfire, will also be
determined to contain a wildfire. If this property holds, then there
is a specific wildfire magnitude threshold, above which the system
will be reliable.  For our purposes, we use the linear planting
function: $p(s, b)=s+b$, as a good approximation to the full
generation function, as it approximately represents real wildfire
signals at their early stages on the background images.

The above definition only considers a single pair of a
signal image stream and a background image stream. Ideally, we would like to verify consistency for
\emph{all} possible background images containing wildfires.  Thus, we
define \emph{global} consistency, as follows:

\begin{newDefinition}[\textbf{Global Consistency}]
  Given a deep neural network 
  $N : \mathbb{R}^{n \times k} \rightarrow \mathbb{R}$, we say that $N$ is
  globally-consistent if for every $x_s \in \mathcal{S}$ and $x_b \in \mathcal{B}$,
 $N$ is $(x_s, x_b)$-locally-consistent.
\end{newDefinition}

We note that the sets $\mathcal{S}$ and $\mathcal{B}$ are not
necessarily finite, and may represent \emph{all} possible wildfire
signal images and \emph{all} possible background images,
respectively. Thus, global consistency is significantly more complex to prove than local consistency. 



\section{Conclusion and Remaining Challenges}



This paper presents a collaboration between academia and
industry, with the goal of verifying an airborne system for wildfire
detection. Our work so far has focused on devising novel kinds of
specifications of interest, which are better suited for this domain
than the specifications commonly supported by academia-oriented
verification tools. Moving forward, we plan to formulate such
properties for the remaining parts of the system, and also to enhance
existing verification engines so that they become sufficiently expressive and
scalable to tackle the networks in question.



\medskip
\noindent
\textbf{Acknowledgements.}  This work was supported by a grant from
the Israel Innovation Authority. The work of Amir was also supported
by a scholarship from the Clore Israel Foundation.


{
\bibliographystyle{abbrv}
\bibliography{references}
}

\newpage
	\appendix
	\renewcommand{\thesection}{\Alph{section}}
	
	\section*{\huge Appendices}
	
	\section{Background: DNNs and their Verification}
	\label{sec:appendix:Background}

\mysubsection{Deep Neural Networks.} A deep neural network
(DNN)~\cite{GoBeCo16} is a computational, directed graph, comprised of
layers. The network computes a 
value, by receiving inputs and propagating them through its layers
until reaching the final (output) layer. These output
values can be interpreted as a classification label or as a regression
value, depending on the kind of network in question. The actual computation
depends on each layer's \emph{type}.  For example, a node $y$ in a
\emph{rectified linear unit} (\emph{ReLU}) layer calculates the value
$y=\relu{}(x)=\max(0,x)$, for the value $x$ of one of the nodes in its
preceding layer. Additional layer types include weighted sum
layers, as well as layers with various non-linear activations. Here,
we focus on \emph{feed-forward} neural networks, i.e., DNNs in which
each layer is connected only to its following layer.

	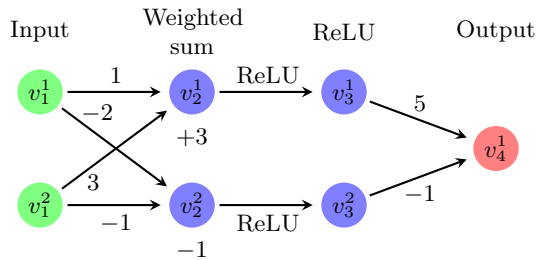
\begin{figure*}[h]
	\centering
		\scalebox{1.0} {
			\def\layersep{2.0cm}
			\begin{tikzpicture}[shorten >=1pt,->,draw=black!50, node 
				distance=\layersep,font=\footnotesize]
				
				\node[input neuron] (I-1) at (0,-1) {$v^1_1$};
				\node[input neuron] (I-2) at (0,-2.5) {$v^2_1$};
				
				\node[hidden neuron] (H-1) at (\layersep,-1) {$v^1_2$};
				\node[hidden neuron] (H-2) at (\layersep,-2.5) {$v^2_2$};
				
				\node[hidden neuron] (H-3) at (2*\layersep,-1) {$v^1_3$};
				\node[hidden neuron] (H-4) at (2*\layersep,-2.5) {$v^2_3$};
				
				\node[output neuron] at (3*\layersep, -1.75) (O-1) 
				{$v^1_4$};
				
				\draw[nnedge] (I-1) --node[above] {$1$} (H-1);
				\draw[nnedge] (I-1) --node[above, pos=0.3] {$\ -2$} (H-2);
				\draw[nnedge] (I-2) --node[below, pos=0.3] {$3$} (H-1);
				\draw[nnedge] (I-2) --node[below] {$-1$} (H-2);
				
				\draw[nnedge] (H-1) --node[above] {$\relu$} (H-3);
				\draw[nnedge] (H-2) --node[below] {$\relu$} (H-4);
				
				\draw[nnedge] (H-3) --node[above] {$5$} (O-1);
				\draw[nnedge] (H-4) --node[below] {$-1$} (O-1);

				\node[below=0.05cm of H-1] (b1) {$+3$};
				\node[below=0.05cm of H-2] (b2) {$-1$};
				
				\node[annot,above of=H-1, node distance=0.8cm] (hl1) 
				{Weighted 
					sum};
				\node[annot,above of=H-3, node distance=0.8cm] (hl2) {ReLU 
				};
				\node[annot,left of=hl1] {Input };
				\node[annot,right of=hl2] {Output };
			\end{tikzpicture}
		}	
	\caption{A toy DNN.}
	\label{fig:toyDnn}
\end{figure*}

Fig.~\ref{fig:toyDnn} depicts a toy DNN. For input $V_1=[1, 3]^T$, the
second layer computes the values $V_2=[13,-6]^T$. In the third layer,
the \relu{} functions are applied, producing $V_3=[13,0]^T$. Finally,
the network's single output value is $V_4=[65]$.
	
\mysubsection{DNN Verification.}  A DNN verification
engine~\cite{KaBaDiJuKo17, GeMiDrTsChVe18, WaPeWhYaJa18,
  LyKoKoWoLiDa20, HuKwWaWu17} receives a DNN $N$, a precondition $P$
that defines a subspace of the network's inputs, and a postcondition
$Q$ that limits the network's output values.  The verification engine
then searches for an input $x_0$ that satisfies
$P(x_0) \wedge Q(N(x_0))$. If such an input exists, the engine returns
\sat and a concrete input that satisfies the constraints; otherwise,
it returns \unsat, indicating that no such input exists.  The
postcondition $Q$ usually encodes the \emph{negation} of the desired
property, and hence a \sat answer indicates that the property is
violated, and that the returned $x_0$ triggers a bug. However, an \unsat
result indicates that the property holds.

For example, suppose we wish to verify that the simple DNN depicted in
Fig.~\ref{fig:toyDnn} always outputs a value strictly larger than
$25$; i.e., for any input $x=\langle v_1^1,v_1^2\rangle$, it holds
that $N(x)=v_4^1 > 25$. This property is encoded as a verification
query by choosing a precondition that does not restrict the input,
i.e., $P=(true)$, and by setting a postcondition $Q=(v_4^1\leq 25)$.
For this verification query, a sound verification engine will return
\sat, alongside a feasible counterexample such as
$x=\langle 1, 0\rangle$, which produces $v_4^1=20 \leq 25$, proving
that the property does not hold for this DNN.
	
In our work, we used \marabou~\cite{KaHuIbHuLaLiShThWuZeDiKoBa19} --- a sound and
complete DNN-verification engine, which has recently been used in a variety of
applications~\cite{AmScKa21, AmZeKaSc22, OsBaKa22, IsBaZhKa22, ReKa22,
  AmCoYeMaHaFaKa22, CoYeAmFaHaKa22, BaKa22, ElGoKa20, ElCoKa22}.
	
\end{document}